\documentclass[useAMS,usenatbib]{mn2e}
\usepackage{graphicx}

\def\gtsima
{\hbox{\raise0.5ex\hbox{$>\lower1.06ex\hbox{$\kern-1.07em{\sim}$}$}}}
\def\ltsima
{\hbox{\raise0.5ex\hbox{$<\lower1.06ex\hbox{$\kern-1.07em{\sim}$}$}}}

\title[The Sgr Nuclear Structure]{The central density cusp of the Sagittarius 
dwarf spheroidal galaxy\thanks{Based on observations made with the
European Southern  Observatory telescopes, using the Wide Field Imager, as part
of the  observing program 65.L-0463. Also based on data obtained from the 
ESO/ST-ECF Science Archive Facility.}}
\author[L. Monaco et al.]{L. Monaco$^{1,2,3}$, M. Bellazzini$^{2}$, 
F.R. Ferraro$^{3}$, E. Pancino$^{2}$
\thanks{E-mail: lmonaco@ts.astro.it; michele.bellazzini@bo.astro.it; 
francesco.ferraro3@unibo.it; elena.pancino@bo.astro.it}\\
$^{1}$INAF - Osservatorio Astronomico di Trieste, Via Tiepolo 11, 34131, 
Trieste, Italy\\
$^{2}$INAF - Osservatorio Astronomico di Bologna, via Ranzani 1, 40127, 
Bologna, Italy\\
$^{3}$Universit\`a di Bologna - Dipartimento di Astronomia, via Ranzani 1, 
40127, Bologna, Italy}

\begin{document}

\date{\today}

\pagerange{\pageref{firstpage}--\pageref{lastpage}} \pubyear{2004}

\maketitle

\label{firstpage}

\begin{abstract} 


We present an analysis of the density profile in the central region of the
Sagittarius dwarf spheroidal galaxy. A strong density enhancement of Sgr stars
is observed. The position of the peak of the detected cusp is
indistinguishable from the center of M54. The photometric properties of the
cusp are fully compatible with those observed in the nuclei of dwarf elliptical
galaxies, indicating that the Sgr dSph would appear as a nucleated galaxy
independently of the presence of M54 at its center.

\end{abstract} 


 
\section{Introduction} 

Dwarf galaxies are considered the building blocks of the hierarchical merging 
process, a fundamental mechanism for the formation of large galaxies. Among
dwarf galaxies, dwarf ellipticals are the most common type of galaxy  in the
nearby universe \cite[][hereafter FB94]{ferguson}.  Therefore, the
comprehension of their structural and evolutionary characteristics  is a major
task of modern cosmology.

A characteristic of many dE is an enhancement of the surface brightness in a 
small central region ({\it nucleus}). Such a feature defines the
sub-class of nucleated dwarf ellipticals (dE,N). 

The observed nuclei have surface brightness profiles similar to globular
clusters, they also share with globulars the same general {\em surface
brightness - absolute magnitude} relation and their luminosity function
overlaps the luminosity range covered by globular clusters
\cite[FB94,][]{durrell,zinneker}.  Hence, the origin of dE nuclei is generally
reconducted to two possible mechanisms, both related to massive star clusters,
e.g. (a) the decay of the orbit of a pre-existing globular toward the tip of
the galactic potential well, driven by dynamical friction, or (b) the {\em in
situ} formation of a giant cluster from gas fallen to the  center of the galaxy
\cite[see][and references therein]{durrell,dp,bassino,vdb}. dE nuclei may also
be related with the Ultracompact Dwarf Galaxies  (UCD), recently discovered in
the Fornax cluster \citep[][]{drink,phi,compact}. In summary, the phenomenon of
dE nucleation is far from being fully understood and it is the subject of
continuous investigation \cite[see, e.g.][]{bbj,stiavelli,miller,ali}.  A local
example would certainly provide a deeper insight on the phenomenon, but the
only well-known case in the Local Group is M~32, a companion of the Andromeda
galaxy whose stellar content may be studied in some detail only with HST
\citep{grillmair}.

Many authors \citep[][hereafter LS00]{bamu,sl95,dacosta,ls00} have suggested 
that the Sagittarius dwarf spheroidal galaxy  \citep[Sgr dSph,][]{s1,igi,s2} is
the relic  of a dE,N and that the  massive globular cluster M~54 is the nucleus
of this galaxy. However, several authors also noted that  Sgr stars of
different age and/or metallicities  are somewhat clustered around M~54
\citep{sdgs1,ls00,steve}. In this paper we  fully resolve, for the first time,
the structure of the overdensity of Sgr stars around M~54, showing that Sgr
would appear nucleated independently of the presence of M~54.

 
\section{The nuclear structure} 

The Sgr stellar content is dominated by a metal-rich population ($[M/H]\simeq
-0.4/-0.6$) with an age of $\sim 4-6$ Gyr \citep[see][and references
therein]{bump}. Therefore, the Color Magnitude Diagram (CMD) of Sgr shows the
typical  features of an old/intermediate-age metal-rich population with a 
clearly defined Red Clump (RC) of He-burning stars and a cool Red Giant Branch
(RGB).  On the other hand, M~54 is a quite old and metal poor globular cluster 
\citep[][hereafter LS00]{bwg,ls00} with an extended blue Horizontal Branch
\citep[HB, see][]{hook}
and a steep RGB. Hence, the evolved stars of the two systems can be easily
discriminated in the CMD and allow us to study the respective spatial
distributions  \citep[see][for examples and discussion]{bhbletter}.

\begin{figure}
\includegraphics[width=84mm]{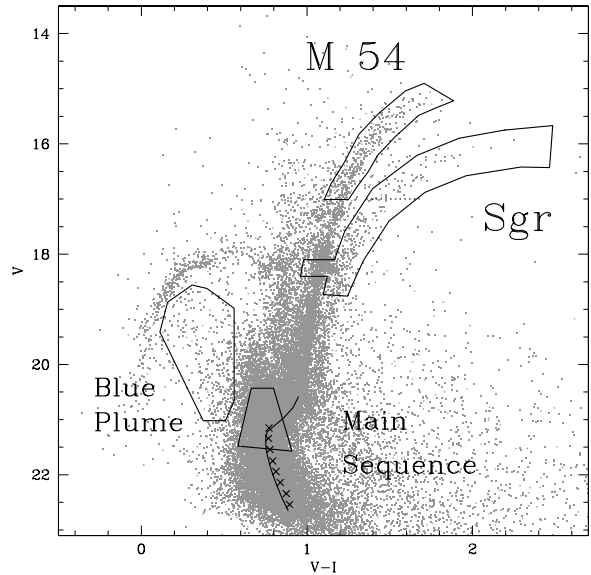}
\caption{
Color-magnitude diagram of the Sgr dSph.  The selection boxes
define the samples of stars  described in the text. In order to better show the
various sequences, only stars in the innermost 4$\arcmin\times$4$\arcmin$
around M~54 are plotted. The M~54 (continuous line) and Sgr (crosses) 
mean ridge lines are also plotted in the main sequence region.
}
\label{uno}
\end{figure}

In the following analysis we will use the photometric catalog of the Sgr dSph
covering a region of about $1^{\circ}\times$1$^{\circ}$ around the globular
cluster M~54 already presented in \citet{bump,bhbletter}.  In order to better
show the various sequences, only stars in the innermost
4$\arcmin\times$4$\arcmin$ around M~54 are plotted in the CMD in Figure
\ref{uno} \citep[the CMD of the entire sample can be found in Figure~1
by][]{bump}. The boxes adopted to select the star samples are shown in 
Fig.~\ref{uno}. Similar boxes have also been defined in \citet[][see their
figure~12]{sdgs1} in order to study the different stellar populations in Sgr.
The reader can also refer to LS00 (see, in particular, figures 14 and 16) for a
field subtracted view of the M~54 and Sgr CMDs.

In comparing the radial distributions of M54 and Sgr we will
rely on Sgr RGB stars having $V\le 17.0$ (hereafter {\it Sgr17}), i.e. with
similar completeness properties and with stars in similar evolutionary phase as
the M54 RGB sample. The larger Sgr sample (from V$\simeq$15.5 down to
V$\simeq$18.7) will be considered in order to trace the Sgr structure out to
the edges of our field with the highest possible signal to noise (see
figure~\ref{norm} below).

We also selected stars in the Blue Plume  (BP) and upper Main Sequence (MS)
regions. We expect the BP sample to be composed by young and, probably, metal rich
Sgr stars. Some blue straggler star belonging to Sgr and/or M~54 may also be
present in  this box. We plotted in the MS region the M~54 (continuous line)
and Sgr (crosses) mean ridge lines. Clearly, both M~54 and Sgr are expected to
contribute to this sample. 

\begin{figure}
\includegraphics[width=84mm]{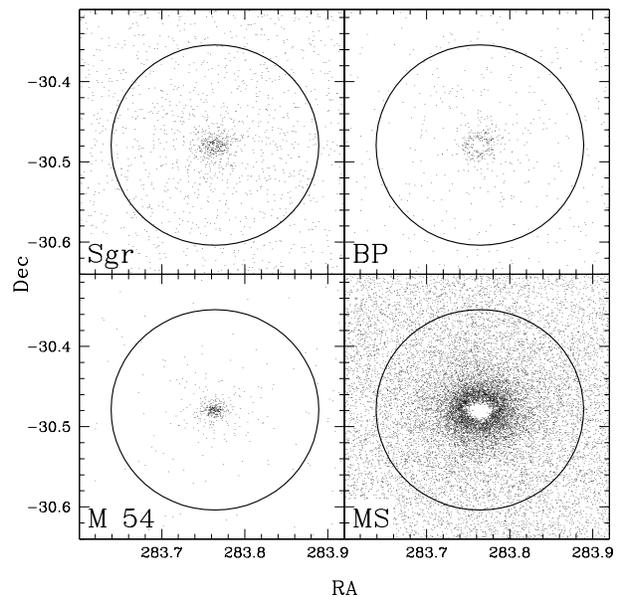}
\caption{
Spatial distribution of the stars in the samples shown in Figure~\ref{uno}.  
In each panel the tidal radius of M~54 ($r^{M54}_t$=7$\arcmin$.5) is marked. 
The {\it ``holes"} in the center of the right-hand panels are due to
incompleteness effects. 
}

\label{due}
\end{figure}

In Figure \ref{due} we show the spatial distribution of stars in the four
samples selected in Fig.~\ref{uno}.  The circle reported in each panel of
Figure \ref{due} represent the tidal radius of M~54
\citep[$r^{M54}_t$=7$\arcmin$.5,][]{trag}. All the samples show a strong
concentration around M~54.  This effect is somewhat obvious in the case of the
M~54 and MS samples (lower panels), which are dominated or at least largely 
contaminated by M~54 stars.   On the other hand, such a sharp concentration
around M~54 is unexpected in the case of the Sgr and BP samples (upper  panels)
where the contamination from M~54 stars is virtually negligible. 

\begin{figure}
\includegraphics[width=84mm]{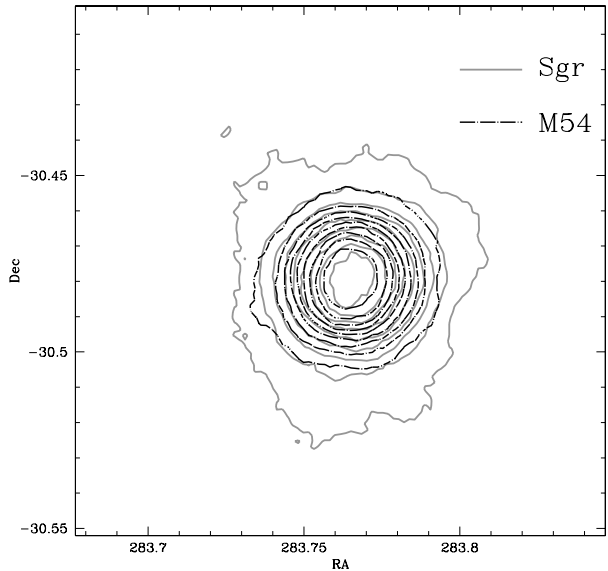}
\caption{M~54 (heavy dot-long dashed curve) and 
Sgr (light grey continuous curve) density contours.}
\label{duebis}
\end{figure}

In Figure \ref{duebis} we show isodensity contours for M~54  (light grey
continuous curves) and Sgr (heavy dotted-dashed curves). Each contour set is
normalized to the respective maximum density  and for each sample we show
isodensity curves ranging from 10\% to the 90\%  of the peak value, in steps of
10\%. The plotted area covers approximately $9\arcmin \times 9\arcmin$.

The M~54 contours show a compact, strongly peaked, symmetric structure, 
typical of a globular cluster.  The surprising feature of Figure \ref{duebis}
is that  Sgr shows a very similar structure centered (at the best of our
resolution, i.e. 6$\arcsec$)  exactly on M~54. 

The scale length of the Sgr overdensity can be best  appreciated by examining
the radial density profile.  Density profiles are obtained by dividing the
surveyed area into quadrants and by computing the density of stars  (star
counts/area) for each quadrant in annuli of variable dimension. For each
annulus, the densities in the quadrants are then averaged to give the final 
density and standard deviation on the density estimates. 

We applied this procedure to stars in the M~54 and in the {\it Sgr17} samples. 
As can be seen from Figure~\ref{uno}, the two boxes select the brightest region
of the RGB, hence the  radial profiles  can be directly compared. In order to
obtain an higher signal in the outer part of the profile,  we also computed the
radial profile for stars in the Sgr global sample\footnote{The Sgr global
sample certainly suffers some degree of contamination by M~54 stars. However,
such a contamination is of the order 3-4\% of the global sample. As can be
realized looking at the upper panel of figure \ref{norm}, the Sgr global sample
and {\it Sgr17} radial profiles are perfectly consistent each other, therefore
the M~54 contamination should be negligible. Such contamination of M~54 stars
do not affect any of the results presented in the paper.}.  Then we {\it
shifted}  the Sgr radial profile in order to match the {\it Sgr17} one in the
region $15\arcsec\ltsima r\ltsima100\arcsec$  (upper panel in Figure
\ref{norm}). The adopted Sgr profile is thus computed  using {\it Sgr17} for
r$\leq25\arcsec$ (i.e. $log(r)\leq1.4$)  and the Sgr sample (normalized to the
former) for r$>25\arcsec$ (lower panel in Figure \ref{norm}). As expected, the
global Sgr sample is heavily incomplete in the innermost region. The effect of
incompleteness is expected to be significantly lower in the {\it Sgr17} sample,
however, to be conservative, we consider the inner points (r$<$16$\arcsec$)
only as lower limits, hence they are plotted with different symbols in
Figure~\ref{norm}.

\begin{figure}
\includegraphics[width=84mm]{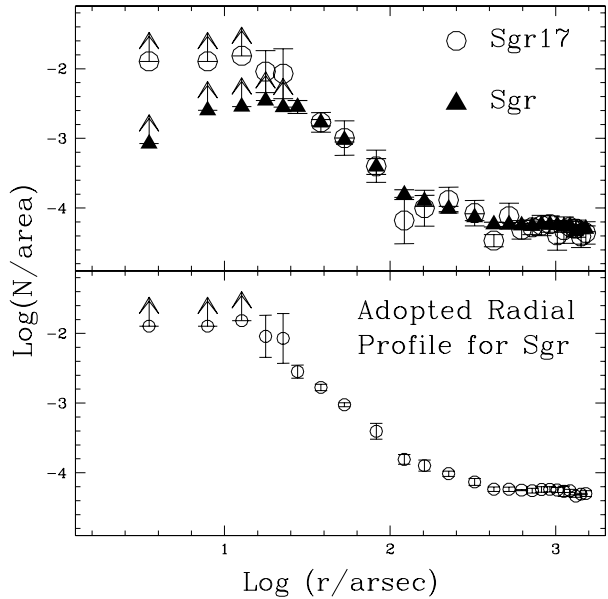}
\caption{Upper panel: Radial profile of stars in the Sgr sample
(filled triangles) and in {\it Sgr17} (empty circles). The Sgr sample 
radial profile have been normalized to match the {\it Sgr17} one.
The adopted composite radial profile for Sgr is shown in the 
lower panel.}
\label{norm}
\end{figure}

\begin{figure}
\includegraphics[width=84mm]{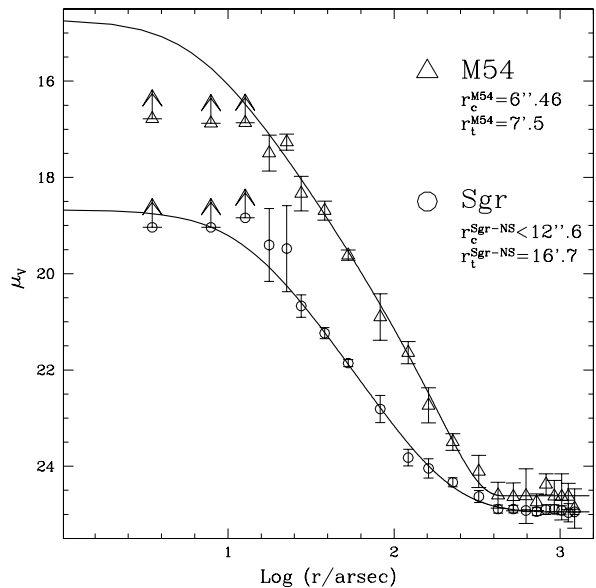}
\caption{Sgr (empty circles) and M~54 (empty triangles) radial profiles. The
continuous curves are King models having the displayed structural parameters.}
\label{pro1}
\end{figure}

In order to compute the radial surface brightness profile we followed two
different approaches: 
\begin{enumerate} 

\item The brightness of each annulus (see above) has been computed as the sum
of the luminosity of all the stars lying in it normalized to the annulus area.
However, as expected, the brightness profile turns out to be quite noisy since
it is affected by large fluctuations due to the small number statistics of few
bright giants. 

\item On the other hand, since the observed number of stars in any post-MS
evolutionary stage is a function of the sampled luminosity and the duration of
the considered evolutionary phase \citep[see][]{rf88}, the star density, 
N/area, can be easily converted into surface brightness by the relation: 
$\mu_{V}=-2.5\times Log(N/area)+ c$, where the value of $c$ depends on the
considered evolutionary phase. Since both the M~54  and  {\it Sgr17} selection
boxes sample approximately the same portion of RGB (see Figure~\ref{uno}), the
same value of the constant {\it ``c"} can be applied to both profiles. We
derived the value of c by fitting the M~54 radial profile with a King
model\footnote{Through the rest of the paper we will call {\it King model}  the
sum of a King model plus a constant component representing the foreground
contamination or the Sgr galaxy field.} having the structural parameters
tabulated in \citet[][]{trag} and normalizing the central surface brightness to
$\mu_{V,0}=14.75$ \citep[][]{trag}.

\end{enumerate}
The radial surface-brightness profile from procedure (ii) turns out to be fully
consistent but less noisy than the one obtained from procedure (i).

The derived radial surface brightness profiles for M~54 and Sgr are plotted in
Figure \ref{pro1} (empty triangles and circles, respectively) together with the
M~54 best fit King model \citep[][upper continuous curve]{trag}.  Note that the surface brightness in
the outer part of the Sgr profile (r$>$10$\arcmin$, $\mu_{V}\simeq24.9$), is
fully consistent with previous determinations by
\citet[][]{s2,musk,ma98,steve}.      

We also fitted the inner Sgr profile (hereafter Nuclear Structure, Sgr-NS) with
a King model (lower continuous curve) having r$^{Sgr-NS}_c$=12$\arcsec$.6
(which should be considered as an upper limit) and r$^{Sgr-NS}_t$=16$\arcmin$.7
as core and tidal radii, respectively. This King model can be used to derive a
lower limit to the Sgr-NS integrated magnitude. It is also important to stress
that a satisfactory fit to the Sgr-NS  profile cannot be obtained using
r$^{M54}_t$ as tidal radius. Thus, as a first result, we conclude that {\it the
Sgr-NS and M~54 do not share the same profile},  the Sgr-NS being a more
extended structure. 

\begin{figure}
\includegraphics[width=84mm]{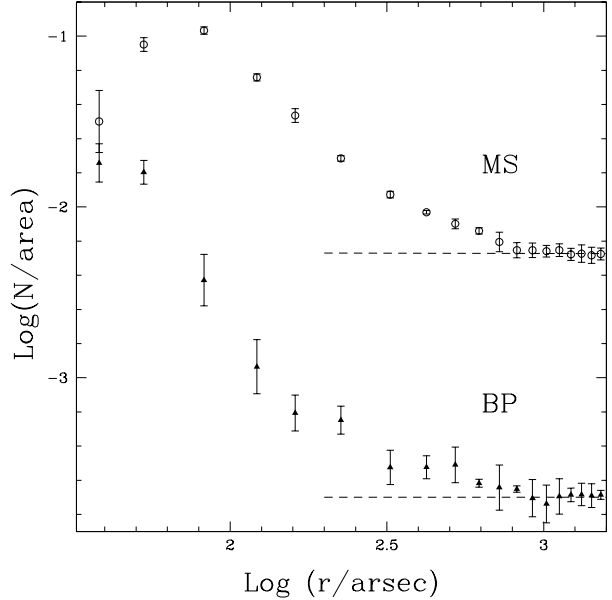}
\caption{Radial profiles for the MS and BP samples selected 
in Figure \ref{uno}. Dashed lines mark the plateau level reached in the most 
external regions of the profiles.}
\label{plms}
\end{figure}

Indeed, the presence of the Sgr-NS and the Sgr field may have corrupted the
r$^{M54}_t$ value derived by \citet{trag}. However, the M~54 radial profile
presented in figure \ref{pro1}, confirms that r$^{M54}_t$=7$\arcmin$.5 is a
reasonable value. In any case, even if r$^{M54}_t$ was actually overestimated,
our conclusions remain unchanged: the Sgr-NS and M54 scale lengths and, hence,
their radial profiles are different. 

In figure \ref{plms} we plotted the radial profiles obtained also for the BP
(filled triangles) and MS (open circles) samples selected in Figure \ref{uno}. 
Once again, we see two significantly peaked structures with respect to a
constant value reached in  the external part of the profiles.  We recall here
that the MS sample is composed by a mix of the M~54 and Sgr stellar populations
while the BP sample is essentially composed by young and possibly metal rich
Sgr stars \citep[see figure \ref{uno} and section 2 above and
LS00,][]{sdgs1,sdgs2}. In the MS profile, the star density excess drops to zero
at about $\sim$13$\arcmin$, thereby it cannot be due only to M~54 stars. The BP
profile, with its peaked shape, confirms the existence of a nuclear structure
in Sgr. The BP profile seems to hold a density excess out to
$\sim$15$\arcmin$,  even if with a low signal which prevents from drawing any
firm conclusion.  The BP in comparison with the MS profile seems to show a
steeper decline.  This somewhat unexpected trend may be, at least partly, due
to the more severe incompleteness effect suffered by the fainter sample, i.e.
the MS sample, in the most central regions.

\begin{figure} 
\includegraphics[width=84mm]{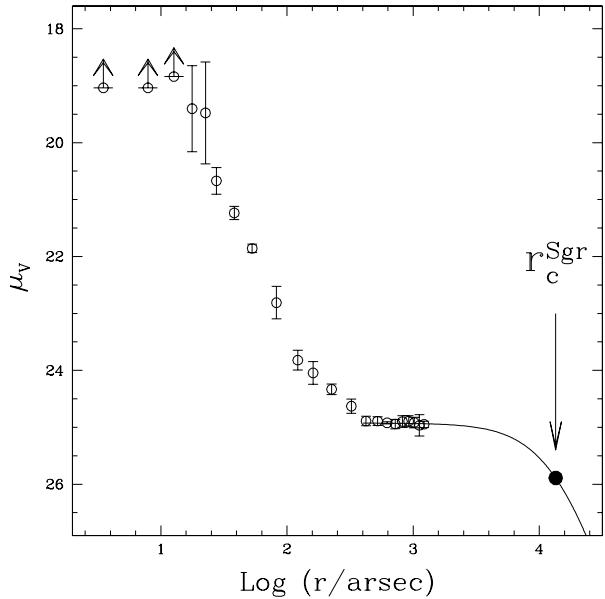} 
\caption{Sgr radial profile.  The continuous curve is a King
model having the structural parameters derived by \citet[][]{steve} for the
main body of Sgr. The filled circle indicates the main body core radius.} 
\label{pro}
\end{figure}

In Figure \ref{pro} the surface brightness profile for the Sgr population
obtained here is combined with the King model obtained by \citet{steve} for the
external part of the Sgr galaxy. The position of the core radius of the main
body of the system (r$_c^{Sgr}$=224$\arcmin$) is also shown in the Figure
(large filled circle).

Figures \ref{pro} and \ref{plms} clearly show the presence of a central cusp
in the Sgr dSph as first suggested by LS00 and more recently by
\citet[][]{steve}.  In particular LS00 (their Figure~19, panel c) found a
density excess of Sgr RGB stars which drops significantly  between 3$\arcmin$ 
to $\sim$12$\arcmin$ from the center of M~54 in agreement with the Sgr profile
in Figure \ref{pro}. They also find (Figure 20, panel a) that stars in the Blue
Plume  (their {\it ``1 Gyr"} sample) show a star density excess which drops to
zero between 12$\arcmin$ and  19$\arcmin$ from the M~54 center, consistently
with our BP profile.

 
\section{The nature of the overdensity of Sgr stars}

In the previous section we demonstrated the presence of a strongly peaked 
structure in the densest region of Sgr. This structure  (Sgr-NS)  is made by
stars  belonging to the typical metal rich stellar population of Sgr (see
Figure \ref{uno})  and hence it is distinct from M~54. Its center, at best of
our resolution,  coincides with that of M~54 (as can be seen from Figure
\ref{duebis}), while the Sgr-NS profile is more extended than the cluster one
(see Figure \ref{pro1}).

The Sgr radial profile shown in Figure \ref{pro} is extremely similar to that
of a nucleated dwarf galaxy  \citep[i.e. the combination of a large scale
profile and a central cusp, see for instance ][]{stiavelli}. In order to test
the compatibility of Sgr-NS  with a dwarf elliptical nucleus, we now put some
limit on its integrated magnitude (M$_V^{Sgr-NS}$).  From the inspection of
Figure~\ref{pro1}, it is  reasonable to assume  $M_V^{Sgr-NS}\gtsima
M_V^{M54}=$-10.01 \citep{harris} as an upper  limit.

Although the Sgr radial profile is certainly affected by severe incompleteness 
(in the innermost region, see Figure~\ref{pro}), it can be used to set a lower
limit to the integrated magnitude of the Sgr central cusp, M$_V^{Sgr-NS}$.  

By integrating the Sgr King model adopted in Figure~\ref{pro1}  (lower curve)
we obtained  $L_{tot,bol}^{Sgr-NS}=2.2\times10^{5}L_\odot$ which corresponds to
M$_V^{Sgr-NS}$=-7.8 by assuming M$_{V\odot}$=4.83 and the appropriate bolometric
correction \citep[see, e.g., Table 5 by][]{buz}.

As in section~2, we performed a consistency check by using star counts.   First
we selected a region in which the incompleteness effects are expected to be
negligible (2$\arcmin<$r$<$4$\arcmin$ from the cusp center). By using the Sgr
King model adopted in Figure \ref{pro1} we estimate  that the annulus  contains
$\sim$16\% of the global cusp luminosity. Then we use the number of HB stars
observed in the considered annulus in order to calibrate the total luminosity.
In the considered annulus we counted 56 HB stars\footnote{We counted the stars 
in a Red Clump selection box and evaluated the foreground contamination by
counting the stars within an adjacent box having the same size \citep[see
also][]{bhbletter}.}. From this number we estimate a total HB population of
N$_{HB}$=56/0.16=350 stars in the central cusp. As already mentioned, the
number of stars observed in any given post-MS evolutionary stage is connected
to the global luminosity and the duration of the phase by the equation:
$N_{J}=B(t)\times t_J\times L$, where N$_J$ and t$_J$ are the number of stars
and the duration of the J evolutionary stage and B(t) is the specific 
evolutionary flux for the considered stellar population
($B(t)=2\times10^{-11}~yr^{-1}L_\odot^{-1}$  for an old stellar population). 
By assuming $t_{HB}=10^8$~yr and N$_{HB}$=350 we easily obtain the global
luminosity: $L_{tot,bol}^{Sgr-NS}=1.8\times10^{5}L_\odot$ which corresponds to
M$_V^{Sgr-NS}$=-7.6 under the same assumption about M$_{V\odot}$ and the
bolometric correction. Thus, the two estimates are in close agreement.
However,  we will consider a conservative global uncertainty of $\pm0.5$~mag
for  our estimate.  

\begin{figure}
\includegraphics[width=84mm]{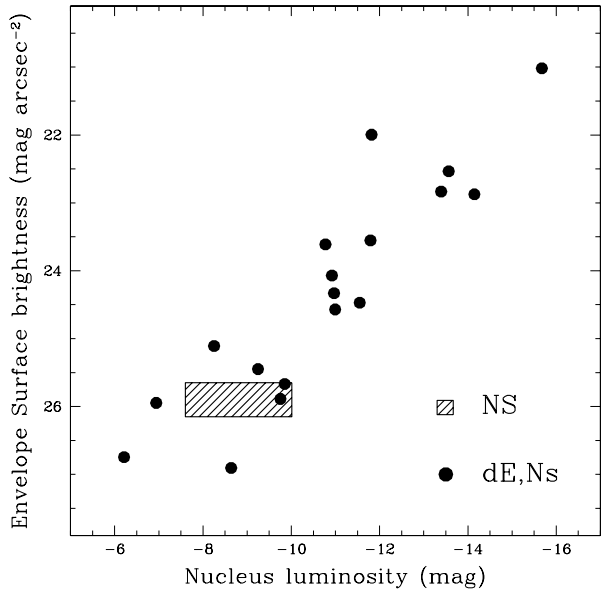}
\caption{The surface brightness of the underlying galaxy is plotted against 
the integrated magnitude of the core for the sample of dE,N presented by 
\citet{compact}. The shaded area in the plane covers the range allowed to 
Sgr-NS by our data.}
\label{bri}
\end{figure}

We can now compare the photometric properties of the Sgr/Sgr-NS system derived
above with those observed in dE,N. In Figure \ref{bri} the Surface Brightness
of the envelope (at half light) is plotted {\it vs} the Nucleus Luminosity for
a sample of dE,N in the Fornax cluster \citep[filled circles,][]{compact} .  The
permitted range of parameter space for Sgr-NS is represented by the shaded
area,  where we assumed as surface brightness of the envelope the one at the
core radius of the main body of Sgr, i.e. $\mu_{V}\simeq25.89$. As can be seen,
in this plane {\it the Sgr-NS  behaves exactly as a nucleus}.

As quoted in Sect.~1., the possibility that M~54 represent the nucleus of Sgr
was already discussed in the  literature. However, as noted by \citet{sl95},
galactic nuclei usually have colors similar to that of the  surrounding galaxy
field \citep{freeman,ferguson,lotz}, even if a few nuclei bluer than the field
do exist \citep{durrell}. At odds with M~54, Sgr-NS has the same color of the
surrounding Sgr field. 

Moreover the Sgr-NS shows a central surface brightness ($\mu_0<$19) more than
6~mag brighter than the central brightness of the main body of Sgr
($\mu_0\simeq$25). This is a pretty typical value for a dE,N. Conversely, if
M~54 is considered as the Sgr nucleus it would have a $\mu_0$ more than 10~mag
brighter than the brightness of the Sgr envelope. Note that none of the 5 dE,N
analyzed by \citet[][]{geha} and only one of the 14 analyzed by
\citet{stiavelli} show such a large difference.    It is therefore safe to
conclude that the structural characteristics of Sgr-NS seem {\it more typical
of a dE nucleus} than those of M~54.

The discovery of the Nuclear Structure presented here clearly demonstrates that
{\it the Sgr galaxy has its own nucleus}. Now the most intriguing property of the
Sgr nucleus is the spatial coincidence with M~54, which poses the main question:
which is the relation between the Sgr-NS and the cluster M~54?

\section{Discussion}

By selecting a sample of Sgr giants in the extensive CMD presented by
\citet[][]{bump}  we have demonstrated the existence of a Nuclear Structure in
the Sgr galaxy.  The radial profile of the  RGB overdensity (Sgr-NS)  is well
reproduced by a King model and it appears different from the profile of M54. 
We considered the characteristics of Sgr-NS  (color, integrated magnitude,
central surface brightness) with respect to  the underlying Sgr field and we
find that Sgr-NS behaves as a typical faint dwarf Elliptical Nucleus. While
M~54 has been claimed to be the Sgr nucleus, the photometric properties of
Sgr-NS are more typical of a dwarf elliptical nucleus than those of M~54. It is
important to remark that {\it the Sgr dSph would appear as a nucleated galaxy
even if M54 would be removed from its center}.

Nevertheless, the spatial coincidence of M~54 and Sgr-NS is impressive  and
needs to be explained. We can envisage three possible frameworks that may be
consistent with the observations. First an {\em in situ} origin can be imagined
for both M54 and Sgr-NS. In this scenario the first generation of (metal-poor)
stars of Sgr formed a bright and compact nucleus at the bottom of the potential
well of the galaxy (M54). The remaining enriched gas formed subsequent
generations of stars following the same density profile, driven by a similar
potential. Hence, the recycled (and chemically enriched) gas accumulated within
and around M54 and subsequently formed the observed Sgr-NS. The difference in
the density profiles of the two systems seems to militate against this
hypothesis. Further interesting clues in this sense may be obtained comparing
the {\em velocity dispersion} profiles of the two RGB samples within
$r_t^{M54}$.

Second, the Sgr-NS can be made by star captured by M~54 from the Sgr field 
\citep[a possibility already suggested by][]{smith98}.  A reliable,
quantitative exploration of this  scenario would require properly designed
dynamical simulations that are clearly beyond the scope of the present
analysis.

Finally, M54 may be an ordinary globular that has been brought to
the central nucleus of Sgr by the dynamical friction. This hypothesis would
naturally explain the spatial coincidence of the cluster and Sgr-NS and the
difference in the density profiles of the two structures that, in this case,
would have different and independent natures. The viability of this scenario
can be verified with simple analytical arguments.

\begin{figure}
\includegraphics[width=84mm]{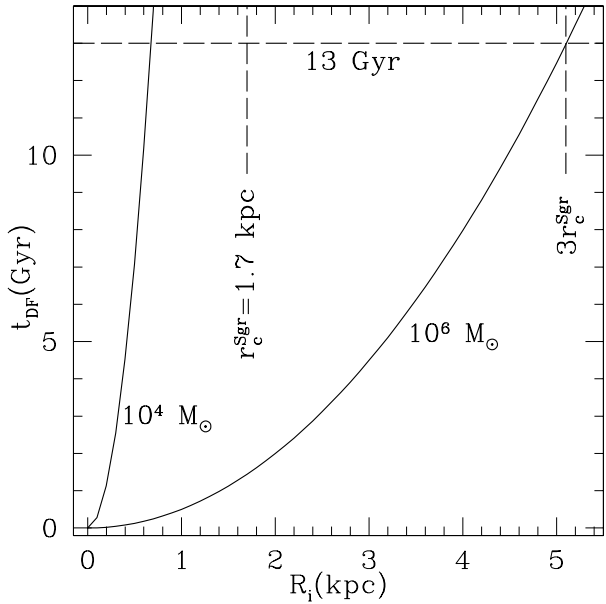}
\caption{Dynamical friction time scale as function of the initial distance from
the Sgr center for M~54 (right continuous curve) and the other Sgr's 
clusters (left continuous curve). An initial velocity of
11.4 km~s$^{-1}$ is assumed.}
\label{dyn}
\end{figure}

The dynamical friction (DF) process is the deceleration experienced by  a
massive object due to the interactions with particles in the surrounding
field. It is an N-body problem which does not admit an exact analytical
solution. However, \citet{bt87} studied the orbital evolution of a massive
object immersed in an isothermal halo. By means of a few assumptions, they were
able to derive a simple formula for the total time, t$_{DF}$,  required for a
body to spiral from an initial radius R$_i$ to the bottom of the  halo
potential well. In Figure \ref{dyn} we plotted t$_{DF}$ as a function of 
R$_i$  for two objects of 10$^6M_{\odot}$ (the M~54 mass), and 10$^4M_{\odot}$
\citep[roughly the mass of the other Sgr's globular clusters,
see][]{pryme,mss}. 

We assumed as initial velocity (relative to the surrounding field) the Sgr
central velocity dispersion as measured by \citet{s2}, $\sigma$=11.4
km~s$^{-1}$. We also assumed the Sgr tidal radius as maximum impact parameter
of the massive body with a halo particle.  We recall here that, assuming a distance
modulus (m-M)$_0$=17.10 \citep{tip}, the core and tidal radii of the Sgr main
body (r$^{Sgr}_c$=224$\arcmin$, r$^{Sgr}_t$=1801$\arcmin$) measured by
\citet{steve} correspond, respectively, to 1.7~kpc and 13.8~kpc.

From the inspection of Figure \ref{dyn}, it is evident that a massive cluster
like  M54 would have been seriously affected by DF in an Hubble time (assumed
to be 13~Gyr) for any initial distance shorter than 3r$^{Sgr}_c$=5.1~kpc from
the galaxy center. Conversely the other Sgr globulars, due to their smaller
masses, would have been affected by DF only for initial distances lower than 
1r$^{Sgr}_c$, while their actual positions are outside r$^{Sgr}_c$ (for
instance, Ter~7 lie at about 3~kpc from M~54).

Hence, the spatial coincidence of NS and M54 can be easily explained if Sgr-NS
was  formed {\it in situ} in the tip of the Sgr potential well and M~54 has
been driven to the galaxy center by dynamical friction.   

\section*{Acknowledgments}

This research is partially supported by the Italian {Ministero 
dell'Universit\`a e della Ricerca Scientifica} (MURST) through the COFIN grant
p.  2002028935, assigned to the project {\em Distance and stellar populations
in the galaxies of the Local Group}. The financial support of ASI is 
acknowledged. We also thank an anonymous referee for useful comments which
significantly improved our paper. Part of the data analysis has been performed
using software developed by P. Montegriffo at the INAF - Osservatorio
Astronomico di Bologna.   This research has made use of NASA's Astrophysics
Data System Abstract Service.


\label{lastpage}

\end{document}